\documentclass[9pt]{article}
\usepackage{Files/spconf}
\usepackage{cite}
\usepackage{amsmath, amssymb, amsfonts, bbold, dsfont, xfrac}
\usepackage{algorithm}
\usepackage{algorithmic}
\usepackage{graphicx}
\usepackage[font=footnotesize]{caption}

\usepackage{textcomp}
\usepackage{hyperref}
\usepackage{xcolor}
\usepackage{comment}
\usepackage{diagbox}
\usepackage{accents}
\usepackage{booktabs}
\usepackage{multirow}
\usepackage{pgfplots}
\pgfplotsset{compat=1.18}
\usepgfplotslibrary{external}
\pgfplotsset{yticklabel style={text width=1.2em,align=right}}
\definecolor{matblue}{rgb}{0, 0.4470, 0.7410}
\definecolor{matred}{rgb}{0.85, 0.3250, 0.0980}
\definecolor{matorange}{rgb}{0.9290, 0.6940, 0.1250}
\definecolor{matviolet}{rgb}{0.4940, 0.1840, 0.5560}
\definecolor{matgreen}{rgb}{0.4660, 0.6740, 0.1880}
\definecolor{matskyblue}{rgb}{0.3010, 0.7450, 0.9330}
\definecolor{matburgundy}{rgb}{0.6350, 0.0780, 0.1840}
\definecolor{matteal}{rgb}{0, .9, .9}
\definecolor{gbblue}{cmyk}{1, .8, .06, .32}

\newcommand{\vect}[1]{\boldsymbol{ #1 }}
\newcommand{\cvar}[2]{\textnormal{CVaR}^{#2}\left[#1 \right]}

\newcommand\authmark[1]{$^{#1}$}

\newtheorem{theorem}{Theorem}
\newtheorem{lemma}{Lemma}

\newtheorem{corollary}{Corollary}

\title{Waterfilling at the Edge:\\Optimal Percentile Resource Allocation via~Risk-Averse Reduction\vspace{-3pt}}

\name{Gokberk Yaylali\authmark{\dag} \qquad Ahmad Ali Khan\authmark{\ddag} \qquad Dionysis Kalogerias\authmark{\dag} \thanks{This work is supported by the NSF under grant CCF 2242215.} \thanks{\textit{Note:} All proofs to the results presented below are deferred to a forthcoming journal submission.}
\vspace{-5pt}}
\address{\authmark{\dag} Department of ECE, Yale University, New Haven, CT 06520 USA\\
\authmark{\ddag} Ericsson R\&D, Ottawa, ON K2K 2V6, Canada \vspace{-13pt}}

\begin{document}
\ninept

\maketitle
\begin{abstract}
\vspace{-5pt}
We address deterministic resource allocation in point-to-point multi-terminal AWGN channels without inter-terminal interference, with particular focus on optimizing quantile transmission rates for cell-edge terminal service. Classical utility-based approaches ---such as minimum rate, sumrate, and proportional fairness--- are either overconservative, or inappropriate, or do not provide a rigorous and/or interpretable foundation for fair rate optimization at the edge. To overcome these challenges, we employ Conditional Value-at-Risk (CVaR), a popular coherent risk measure, and establish its equivalence with the sum-least-$\alpha$th-quantile (SL$\alpha$Q) utility. This connection enables an exact convex reformulation of the SL$\alpha$Q maximization problem, facilitating analytical tractability and precise and interpretable control over cell-edge terminal performance. Utilizing Lagrangian duality, we provide (for the first time) parameterized closed-form solutions for the optimal resource policy ---which is of waterfilling-type---, as well as the associated (auxiliary) Value-at-Risk variable. We further develop a novel inexact  dual subgradient descent algorithm of minimal complexity to determine globally optimal resource policies, and we rigorously establish its convergence. The resulting \textit{edge waterfilling algorithm} iteratively and efficiently allocates resources while explicitly ensuring transmission rate fairness across (cell-edge) terminals. Several (even large-scale) numerical experiments validate the effectiveness of the proposed method for enabling robust quantile rate optimization at the edge.
\end{abstract}
\vspace{-4pt}

\vspace{-7pt}
\section{Introduction}
\vspace{-7pt}

\label{sec:introduction}

We study the classical deterministic resource allocation problem in a point-to-point multi-terminal AWGN communication setting with no inter-terminal interference. As wireless networks evolve to support next-generation applications, efficient and equitable resource allocation across terminals becomes increasingly critical. The inherently variable nature of wireless propagation ---especially under noise-limited scenarios--- causes significant disparities in achievable rate, power efficiency, and throughput across terminals, particularly penalizing cell-edge users. Emerging wireless standards have introduced constraints and performance targets on particular quantile transmission rates \cite{5g_targets_2022, 3gpp_reqs_2024}, yet the direct formulation and solution of quantile rate optimization problems remain largely unexplored. Existing efforts often address the issue indirectly—either through physical-layer strategies \cite{chang_fractional_2016, mankar_hetero_2018, lopez_densify_2015} that avoid explicit signal processing methods, or through signal-processing-based techniques such as beamforming and power control \cite{khan_percentile_2024}, typically grounded in classical network utility maximization paradigms such as minimum rate \cite{luo_sumrate_2008, naghsh_minrate_2019, razaviyayn_minrate_2011}, sumrate \cite{luo_sumrate_2008, evangelista_sumrate_2019}, or proportional fairness \cite{khan_propfair_2018, shen_propfair_2018, shi_propfair_2011, zhang_wsr_2011}.

Minimum-rate maximization seeks to uplift the worst-off users, but proves overconservative and inadequate in large-scale networks where the minimum achievable rate converges to zero with increasing terminal count in the network \cite{ghazanfari_minrate_2020}. In contrast, sumrate maximization disproportionately favors well-positioned users, offering poor service to cell-edge terminals due to opportunistic behavior \cite{ribeiro_optimal_2012, rahman_unfair_2009}. Proportional fairness attempts to strike a compromise, yet it remains an ad-hoc and noninterpretable approach (at least with respect to cell-edge user allocation), and with no formal link to quantile-based performance objectives \cite{khan_percentile_2024}. Recently, \cite{khan_percentile_2024} introduced a numerical scheme for quantile rate maximization via a \textit{sum-least-$\alpha^\text{th}$-quantile (SL$\alpha$Q) utility} ---referred to as sum-least-$q^\text{th}$-percentile (SLqP) in \cite{khan_percentile_2024}--- relying on convex optimization tools such as CVX. Although effective, such generic solver-based approaches are often numerically inefficient at large scales, also lacking a rigorous analytical framework for characterizing performance for the particular problem at hand, where cell-edge user behavior is most critical.

\textit{Risk-averse formulations} offer a promising perspective in the context of resource allocation for wireless networks and relevant applications. Coherent risk measures ---particularly the Conditional Value-at-Risk (CVaR) herein--- are well-established tools for capturing (tail-restricted) risk in stochastic systems, deeply rooted in Finance and Operations Research \cite{shapiro_lectures_2021, rockafellar_optimization_2000}, and have gained increasing relevance in wireless communications for modeling rate reliability under uncertainty \cite{vu_ultra_reliable_2018, li_efficient_2021, bennis_ultrareliable_2018, kalogerias_fast_2022}. Prior work has applied CVaR in a stochastic fading channel setting for robust resource allocation under fading-induced uncertainty \cite{yaylali_robust_2023, yaylali_stochastic_2024, yaylali_distributionally_2024}, showing the inherent ability of CVaR to effectively capture and mitigate the risk of deep fade events.

In this paper, we establish and rely on a fundamental equivalence between deterministic SL$\alpha$Q-based rate optimization in a multi-terminal AWGN channel with varying noise levels, and CVaR-based resource allocation for a single-terminal stochastic fading channel. By interpreting noise variances across terminals as the inverse values of a discrete random fading channel, we uncover a one-to-one correspondence between the two formulations. This allows us to reformulate SL$\alpha$Q maximization using the CVaR, resulting in a convex and analytically tractable resource allocation problem. 

\textit{\textbf{Contributions:}} We first present the reformulation of the SL$\alpha$Q problem via the CVaR, by establishing equivalence with the SL$\alpha$Q utility. Using Lagrangian duality, we then derive closed-form characterizations of the optimal resource allocation and auxiliary CVaR variables. To the best of our knowledge, we provide the first closed-form derivation of an optimal policy for the SL$\alpha$Q problem ---and of waterfilling type---, thereby bridging the gap between tractable implementation and risk-sensitive/fair design. We also develop an inexact subgradient-based dual descent method to recover globally optimal policies, together with a comprehensive convergence analysis, ensuring convergence, superior numerical efficiency and strong scalability, even for very large networks. The resulting \textit{edge waterfilling algorithm}, suitable for real-world applications, is supported by numerical experiments readily demonstrating its effectiveness. 

\vspace{-8pt}
\section{System Model}
\vspace{-5pt}
\label{sec:system_model}

We consider a $N_U$-terminal parallel point-to-point communication channel with no cross-interference and deterministic fading. Also, without loss of generality, we assume constant unity channels gains, i.e., no fading effect on the transmission links (but variable noise variances). The interference-free model is motivated due to simplicity and analytical tractability. Despite the fact that our setting is simplified, the no-interference scenario actually applies to a wide range of real-world communication scenarios, including free-space links (e.g., base station communications), satellite networks, and deep-space communications, to name a few. Cross-interference models remain an engaging perspective to investigate in the future. The transmission rate of terminal $i \in [N_U]$ is modeled as
\begin{equation}
\label{eqn:rate_function}
r_i(p_i, \sigma_i^2) = \log\left( 1 + \frac{p_i}{\sigma_i^2} \right),
\end{equation}
where $p_i \geq 0$ and $\sigma_i^2$ are the resource (power) and the noise variance of the corresponding link $i \in [N_U]$, respectively. To enhance throughput performance for cell-edge users, \cite{khan_percentile_2024} introduces the problem of maximization of the sum of least $\alpha$-quantile (SL$\alpha$Q) transmission rates, i.e.,
\begin{equation}
\label{eqn:khan_problem_formulation}
\begin{aligned}
P_{N}^* = \underset{\vect{p} \succeq \vect{0}}{\textnormal{maximize}} \quad & f_{N_\alpha}(\vect{r}(\vect{p}, \vect{\sigma}^2))\\
\textnormal{subject\ to} \quad & 0 \leq P_0 - \| \vect{p} \|_1,
\end{aligned}
\end{equation}
where $\| \vect{p} \|_1 = \sum_{i=1}^{N_U} p_i$, the utility function $f_{N_\alpha}(\cdot)$ is defined as the sum of the smallest $N_\alpha$ components of a vector $\vect{x}$, i.e., $f_{N_\alpha}(\vect{x}) = \sum_{i = 1}^{N_\alpha} x_i^\uparrow$, where $\vect{x}^\uparrow$ denotes the vector obtained by sorting the elements of $\vect{x}$ in ascending order, $N_\alpha$ is number corresponding to the $\alpha$-quantile user across terminals, i.e., $N_\alpha = \min\{ n \in \mathbb{Z}_{++} \ |\  n \geq \alpha N_U \}=\lceil \alpha N_U \rceil$, $P_0$ is the total power budget, and $\vect{r}$ and $\vect{p}$ are the rate and resource vectors consisting of the corresponding rates and powers $r_i, p_i\ i \in [N_U]$, respectively. We utilize CVaR to facilitate solving the SL$\alpha$Q problem in \eqref{eqn:khan_problem_formulation}, where CVaR is defined for a random integrable \textit{cost} $Z$ as
\begin{equation}
\label{eqn:cvar_definition}
\cvar{Z}{\alpha} = \inf_{t \in \mathbb{R}} \left\{ t + \alpha^{-1} \mathbb{E}\left[ (Z  - t)_+ \right] \right\},
\end{equation}
where $\alpha \in (0, 1]$ is the corresponding confidence level, and $(\cdot)_+ = \max\{ \cdot, 0 \}$. CVaR is a coherent risk measure \cite{shapiro_lectures_2021}, quantifying the expected cost within the upper $\alpha$-quantile of the underlying distribution of $Z$. When $Z$ represents a \textit{reward} (as in our rate \textit{maximization} setting), CVaR must be reflected as
\begin{equation}
\label{eqn:cvar_sup_definition}
-\cvar{-Z}{\alpha} = \sup_{t \in \mathbb{R}} \left\{ t - \alpha^{-1} \mathbb{E}\left[ (t - Z)_+ \right] \right\},
\end{equation}
measuring expected rewards within lower $\alpha$-quantiles.

Even though the original SL$\alpha$Q problem in \eqref{eqn:khan_problem_formulation} is entirely deterministic, the objective function $f_{N_\alpha}(\cdot)$ closely resembles the structure of CVaR. To see this, we introduce a pseudo-stochastic interpretation of the problem; this will enable a more analytically tractable treatment of the original problem. Specifically, we reinterpret the multi-terminal deterministic setting as a single-terminal communication channel with random noise variance $\sigma^2$ taking values in the finite set $\Sigma = \left\{ \sigma_1^2, \cdots, \sigma_{N_U}^2 \right\}$. This construction bridges the deterministic SL$\alpha$Q problem with the stochastic framework previously studied in \cite{yaylali_robust_2023, yaylali_stochastic_2024}, where continuous fading distributions were primarily considered. In fact, the function $f_{N_\alpha}(\vect{r}(\vect{p}, \vect{\sigma}^2))$ may be interpreted as the sum of the lowest $\alpha$-quantile values of the random rate function $r(p(\sigma^2), \sigma^2)$, induced by the (fictitious) random variable $\sigma^2 \in \Sigma$, and where $p(\sigma^2)$ is now the resource \textit{policy}. 

\begin{lemma}[SL$\alpha$Q via CVaR]
\label{lemma:generalization}
For discretized selections of $\alpha$, i.e., $\Bar{\alpha} \in \left\{ \sfrac{k}{N_U} \ \big|\ k \in [N_U] \right\}$ such that $\Bar{\alpha} N_U = N_{\Bar{\alpha}}$, the objective in \eqref{eqn:khan_problem_formulation} admits an exact representation in terms of CVaR, i.e.,
\begin{equation}
\label{eqn:slaq_cvar_equiv}
\frac{1}{N_{\Bar{\alpha}}} f_{N_{\Bar{\alpha}}} (\vect{r}(\vect{p}, \vect{\sigma}^2) = -\cvar{-r(p(\sigma^2), \sigma^2)}{\Bar{\alpha}}.
\end{equation}
\end{lemma}
Capitalizing on Lemma~\ref{lemma:generalization} and the CVaR definition \eqref{eqn:cvar_sup_definition}, we may reformulate problem \eqref{eqn:khan_problem_formulation} for a single-terminal channel with a CVaR objective quantifying the noise variance-induced risk in the transmission rate, i.e.,
\begin{equation}
\label{eqn:cvar_problem_formulation}
\begin{aligned}
P_{R}^* = \underset{p \geq 0,\ t \in \mathbb{R}}{\textnormal{maximize}} \quad & t - \alpha^{-1} \mathbb{E}\left[ ( t - r(p(\sigma^2), \sigma^2))_+ \right]\\
\textnormal{subject\ to} \quad & 0 \leq \Bar{P}_0 - \mathbb{E}[p(\sigma^2)],
\end{aligned}
\end{equation}
where $p(\sigma^2)$ is the resource \textit{policy} admitting the discrete random variable $\sigma^2 \in \Sigma$, $t \in \mathbb{R}$ is the auxiliary Value-at-Risk (VaR) variable \cite{rockafellar_variational_1998}, and $\Bar{P}_0 = \sfrac{P_0}{N_U}$ is the \textit{average} power budget. Problem \eqref{eqn:cvar_problem_formulation} is convex due to the coherence of CVaR. Further, \eqref{eqn:cvar_problem_formulation} exhibits strong duality under certain constraint qualifications (such as Slater's condition, trivially satisfied here), facilitating the utilization of the dual problem within the Lagrangian duality framework. The Lagrangian of the primal problem \eqref{eqn:cvar_problem_formulation} may be expressed as \begin{multline}
\label{eqn:lagrangian_function}
\mathcal{L}(p(\sigma^2), t, \mu) \triangleq t - \alpha^{-1} \mathbb{E}\left[ (t - r(p(\sigma^2), \sigma^2))_+ \right] \\
+ \mu \left( \Bar{P}_0 - \mathbb{E}\left[ p(\sigma^2) \right] \right),
\end{multline}
where $\mu \geq 0$ is a Lagrangian multiplier ---the dual variable associated with the power budget constraint in \eqref{eqn:cvar_problem_formulation}. The dual function is subsequently defined as the maximization of the Lagrangian with respect to the primal variables, i.e.,
\begin{equation}
\label{eqn:dual_function}
q(\mu) \triangleq \sup_{p \geq 0, t \in \mathbb{R}} \mathcal{L}(p(\sigma^2), t, \mu),
\end{equation}
which is immediately followed by the dual problem as a minimization of the dual function over the dual variable, i.e.,
\begin{equation}
\label{eqn:dual_problem}
\begin{aligned}
D_R^* &\triangleq \inf_{\mu \geq 0} q(\mu)
= \inf_{\mu \geq 0} \sup_{p \geq 0, t \in \mathbb{R}} \mathcal{L}(p(\sigma^2), t, \mu).
\end{aligned}
\end{equation}
We hereafter rely on the risk-averse resource allocation framework laid-out above to  explicitly characterize the dual-parameterized primal solutions pertaining to the minimax formulation in \eqref{eqn:dual_problem}, particularly the policy $p$, as follows.
\vspace{-6pt}
\section{Edge Waterfilling}
\vspace{-5pt}
\label{sec:RA_resource_allocation}

We re-organize the terms in the dual function and decompose into subproblems to enable tractable analysis. Generically leveraging the interchangeability principle for expectation \cite[Theorem~9.108]{shapiro_lectures_2021}, we rewrite the dual function as
\begin{equation*}
q(\mu) = \mu \Bar{P}_0 + \sup_{t \in \mathbb{R}}\ t + \mathbb{E} \bigg[\sup_{p \geq 0}\ -\alpha^{-1} (t - r(p, \sigma^2))_+ - \mu p \bigg].
\end{equation*}
This representation enables solving the problem by addressing the inner subproblems in the expectation above in sequence.

\subsection{Optimal Resource Policy}

We address the subproblem with respect to resource policy $p$ using \eqref{eqn:rate_function}, which is expressed as
\begin{equation}
\label{eqn:p_subproblem}
\sup_{p \geq 0}\ -\alpha^{-1} \left(t - \log\left( 1 + \frac{p}{\sigma^2} \right) \right)_+ - \mu p.
\end{equation}
The following central result characterizes the parameterized optimal solution of \eqref{eqn:p_subproblem} explicitly in closed form.
\begin{theorem}[Optimal Resource Policy]
A parameterized solution to the subproblem \eqref{eqn:p_subproblem} may be expressed as
\begin{equation}
\label{eqn:p_solution}
p^*(\sigma^2) = \min\left\{ \left( \frac{1}{\mu \alpha} - \sigma^2 \right)_+, \sigma^2\left(e^t - 1 \right) \right\},
\end{equation}
where $\sigma^2 \in \Sigma :=  \left\{ \sigma_1^2, \cdots, \sigma_{N_U}^2 \right\}$.
\label{thr:p_theorem}
\end{theorem}
The proof is omitted, but can be reconstructed as in \cite{yaylali_robust_2023, yaylali_stochastic_2024}.

\subsection{Optimal Value-at-Risk Levels}

To address the subproblem with respect to the auxiliary Value-at-Risk (VaR) \cite{rockafellar_variational_1998} variable $t$, we recall the dual function and express the subproblem as
\begin{equation}
\label{eqn:t_subproblem}
\sup_{t \in \mathbb{R}}\  t - \mathbb{E}\left[ \alpha^{-1} \left( t - \log\left( 1 + \frac{p^*}{\sigma^2} \right) \right)_+ + \mu p^* \right],
\end{equation}
by substituting the optimal $p^*$. We present a (quasi-)closed-form solution for the optimal VaR level $t^*$ in the next result.
\begin{theorem}[Optimal Value-at-Risk]
\label{theorem:t_solution}
An optimal solution for \eqref{eqn:t_subproblem} $t^* \geq 0$ exists and satisfies the equation
\begin{equation}
F_{\sigma^2}\left( \frac{e^{-t^*}}{\mu \alpha} \right) - \frac{\mu \alpha}{e^{-t^*}} \mathbb{E}\left[ \sigma^2 \cdot \mathds{1}_{\left\{ \sigma^2 \leq \frac{e^{-t^*}}{\mu \alpha} \right\}} \right] = 1 - \alpha,
\end{equation}
where $F_{\sigma^2}$ is the cdf of random variable $\sigma^2$.
\end{theorem}
Since the fictitious discrete distribution $F_{\sigma^2}(\cdot)$ is readily available in closed-form, the solution $t^*$ can be efficiently evaluated via numerical methods, in particular bisection within a tolerance $\epsilon$. 

\begin{algorithm}[tbp]
\centering
\begin{algorithmic}
\STATE Choose initial values $t^{(0)}, \mu^{(0)}$, $\epsilon$, $\gamma$.
\FOR{$n = 1$ \textbf{to} Process End}
\STATE \textit{\# Primal Variables}
\STATE $\vect{\to}$ Set $t^*(\cdot)$ using Theorem~\eqref{theorem:t_solution} (inexactly at tolerance $\epsilon$).
\STATE $\vect{\to}$ Set $p^*\left( \cdot \right)$ using \eqref{eqn:p_solution}, for all $\sigma^2 \in \Sigma$.
\STATE \textit{\# Dual Variables}
\STATE $\vect{\to}$ Update $\mu^{(n)}$ using \eqref{eqn:dual_descent} and \eqref{eqn:mu_subgradient}.
\ENDFOR
\end{algorithmic}
\caption{Edge Waterfilling (EW)}
\label{alg:edge_waterfilling}
\end{algorithm}

\subsection{Dual Descent}

We restate the dual problem by substituting the optimal resource policy $p^*$ and the optimal VaR variable $t^*$ as
\begin{equation*}
D^* = \inf_{\mu \geq 0}\ t^* - \alpha^{-1} \mathbb{E}\left[ (t^* - r(p^*,\cdot))_+ \right] + \mu \left( \Bar{P}_0 - \mathbb{E}\left[ p^* \right] \right).
\end{equation*}
Utilizing the strict feasibility of \eqref{eqn:cvar_problem_formulation}, optimal dual variables are compactly supported, as follows.
\begin{lemma}[Set of Dual Solutions]
\label{lemma:mu_optimal_set}
Every optimal dual variable lies in the interval
\begin{align*}
\mathcal{D} = \left\{ \mu \geq 0 \ \bigg|\ \mu \leq \frac{q(\Tilde{\mu}) - \Bar{t} + \alpha^{-1} \mathbb{E}[(\Bar{t} - r(\Bar{p},\cdot))_+] }{\Bar{P}_0 - \mathbb{E}[\Bar{p}]}\triangleq K_\mathcal{D}\right\},
\end{align*}
for an arbitrary $\Tilde{\mu} \geq 0$, and where $(\Bar{p}, \Bar{t})$ is any Slater vector for \eqref{eqn:cvar_problem_formulation}.
\end{lemma}
We exploit the power constraint violation to devise a subgradient descent scheme tailored for dual variable optimization, i.e.,
\begin{equation}
\label{eqn:dual_descent}
\mu^{(n)} = \mathcal{P}_{\mathcal{D}}\left( \mu^{(n-1)} - \gamma g_\mu(\mu^{(n-1)}) \right),
\end{equation}
where $\gamma$ is stepsize, a subgradient $g_\mu(\cdot)$ of $q(\mu)$ at $\mu$ is
\begin{equation}
\label{eqn:mu_subgradient}
g_\mu(\mu) = \Bar{P}_0 - \mathbb{E}\left[ p^*(\mu, \cdot) \right],
\end{equation}
and $\mathcal{P}_{\mathcal{D}}(\cdot)$ is the projection onto the set $\mathcal{D}$ of optimal dual solutions. The proposed \textit{edge waterfilling} scheme to obtain the globally optimal resource policy is presented in Algorithm~\ref{alg:edge_waterfilling}, where we note that $t^*$ is  \textit{realized inexactly by an available estimate} $\hat{t}^*$, at tolerance $\epsilon$.

For such a tolerance $\epsilon$ in optimality violation of the VaR variable $t$ due to numerical computations (bisection), i.e., $|\hat{t}^* - t^* | \leq \epsilon$, the dual value optimality violation at the time-average dual iterate such that $\Bar{\mu}^{(n)} = \frac{1}{n}\sum_{i=0}^{n-1} \mu^{(i)}$ can be favorably bounded, as elaborated in the following result.
\begin{theorem}[Convergence Bound I]
\label{theorem:mu_opt_error} There are numbers $K_{\mathrm{VaR}}$ and $K_g$ such that, at every iteration index $n\ge0$,
\begin{equation*}
q(\Bar{\mu}^{(n)}) - P_R^* \leq \frac{| \mu^{(0)} - \mu^* |^2}{2 n \gamma} + 2 \epsilon \Bar{\sigma}^2 K_{\mathrm{VaR}} K_\mathcal{D} + \frac{\gamma K_{g}^2}{2},
\end{equation*}
where $\Bar{\sigma}^2 = \frac{1}{N_U} \sum_{i=1}^{N_U} \sigma_i^2$. 
\end{theorem}

Particular selections of tolerance $\epsilon$ and stepsize $\gamma$ yields a decaying bound on the optimality violation, presented in Corollary~\ref{corollary:tolerance_stepsize}.
\begin{corollary}[Convergence Bound II]
\label{corollary:tolerance_stepsize}
For any fixed iteration budget $T$, tolerance $\epsilon = \sfrac{1}{\sqrt{T}}$ and stepsize $\gamma = \sfrac{1}{\sqrt{T}}$, it is true that
\begin{equation*}
q(\Bar{\mu}^{(T)}) - P^* \leq \frac{| \mu^{(0)} - \mu^* |^2 + 4 \Bar{\sigma}^2 K_{\mathrm{VaR}} K_\mathcal{D} + K_{g}^2}{2 \sqrt{T}}.
\end{equation*}
\end{corollary}

 
By exploiting bisection, the per-iteration complexity of the proposed algorithm is of order $\mathcal{O}(N_U \log{\frac{1}{\epsilon}})$, leading to a favorable total complexity of order $\mathcal{O} ( N_U T \log{\sqrt{T}} )$ over $T$ iterations as per Corollary~\ref{corollary:tolerance_stepsize}, which is notably linear relative to the network size.

\vspace{-8pt}
\section{Performance Evaluation}
\vspace{-6pt}

\label{sec:performance_eval}

\begin{figure}[t]
\centering
\includegraphics[width=\linewidth]{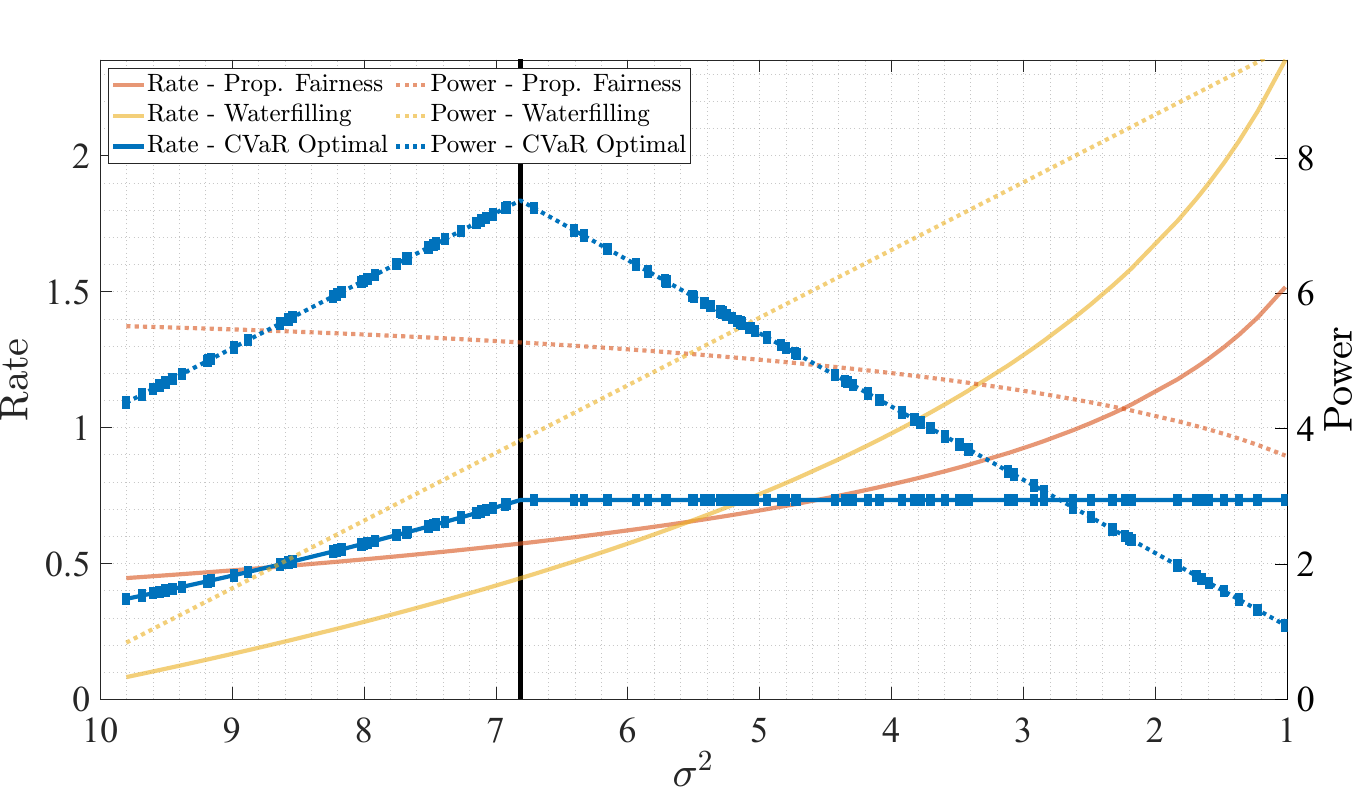}
\caption{Achieved rates and CVaR-optimal policies of terminals distributed by $\mathcal{U}[1,10]$, where $N_U = 100$, $\alpha = 0.75$, $\Bar{P}_0 = 5$, $\gamma = 10^{-3}$, $\epsilon = 10^{-6}$.}
\label{fig:n100_rates}
\end{figure}

\begin{figure}[t]
\centering
\includegraphics[width=\linewidth]{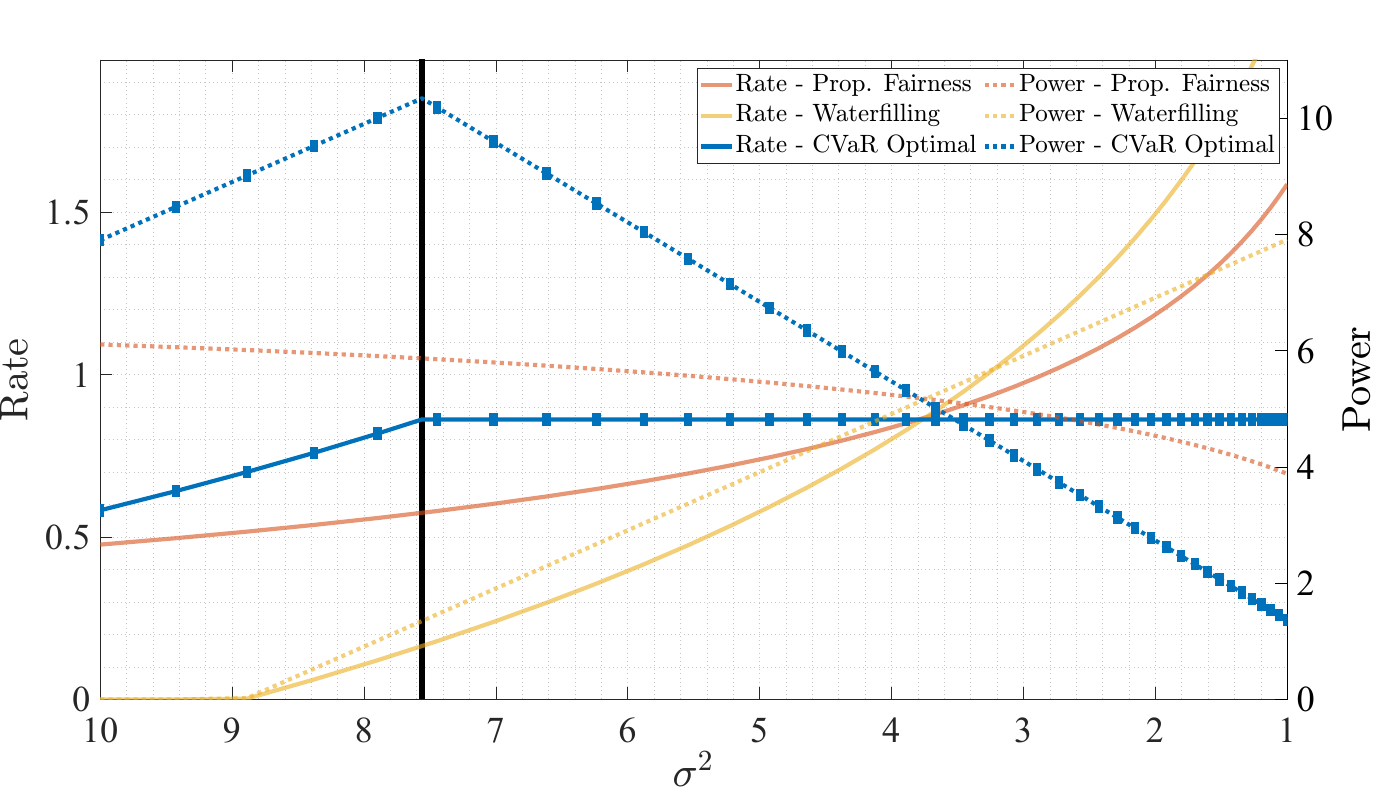}
\caption{Achieved rates and CVaR-optimal policies of logspaced terminals in $[1, 10]$, where $N_U = 40$, $\alpha = 0.5$, $\Bar{P}_0 = 5$, $\gamma = 10^{-3}$, $\epsilon = 10^{-6}$.}
\label{fig:n40_rates}
\end{figure}

We now confirm the effectiveness of the proposed edge waterfilling algorithm by comparing it with benchmark schemes of classical waterfilling and proportional fairness on two transmission scenarios. We first observe that the CVaR-optimal policies introduce a distinct statistical threshold on admissible transmission rates, determined by the optimal VaR variable $t^*$. In particular, the risk-averse nature of the CVaR framework enforces a consistent transmission rate for proximal terminals with low noise variances, promoting fairness by avoiding resource over-concentration. Cell-edge terminals with higher noise variances receive power allocations yielding transmission rates decaying inversely with their variance levels as per the waterfilling-type part of the solution, shown in Figs.~\ref{fig:n100_rates} and \ref{fig:n40_rates}. On the other extent, proximal terminals within a certain proximity ---subject to noise variance--- observe constant transmission rate which is indeed regulated by VaR variable $t^*$. This behavior reflects a statistical cutoff, which is derived from \eqref{eqn:p_solution} as $\hat{\sigma}^2 = \sfrac{e^{-t^*}}{\hat{\mu} \alpha}$ (as shown in Figs.~\ref{fig:n100_rates} and \ref{fig:n40_rates} as vertical line) where $\hat{\mu}$ is the dual variable at convergence. These results are also consistent with Remark 4 in \cite{khan_percentile_2024} which proves that the optimal solution to (\ref{eqn:cvar_problem_formulation}) must display the aforementioned equality of transmission rates for the $N_U-N_{\alpha}+1$ terminals with smallest noise variances. Another (expected) remark is that Edge Waterfilling achieves higher SL$\alpha$Q values, shown in Table~\ref{table:slaq}, proving the effectiveness and fairness among the worst $\alpha$-quantile, as intended. The overall gain is attained without degrading performance at proximal terminals, preserving overall allocation balance.

\begin{figure}[t]
\centering
\includegraphics[width=\linewidth]{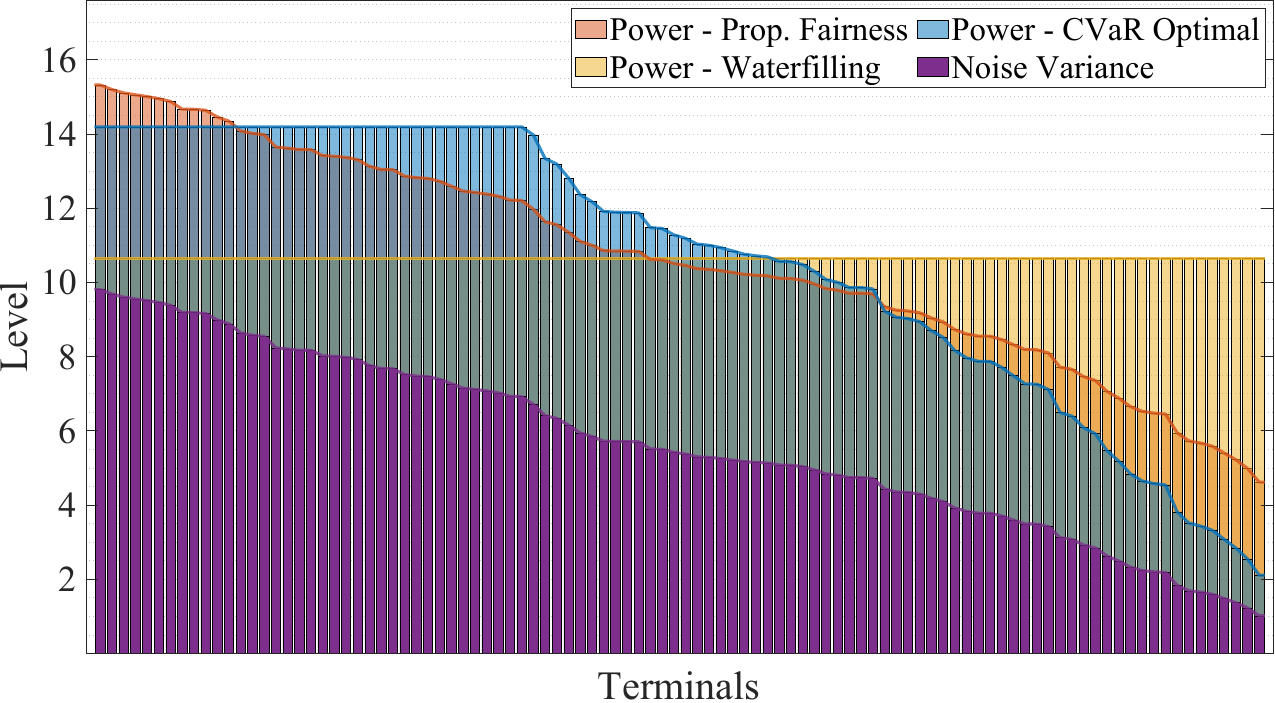}
\caption{Collective levels of allocated policies of terminals distributed by $\mathcal{U}[1,10]$ under Edge Waterfilling and the benchmark schemes.}
\label{fig:n100_bar}
\end{figure}

\begin{figure}[t]
\centering
\includegraphics[width=\linewidth]{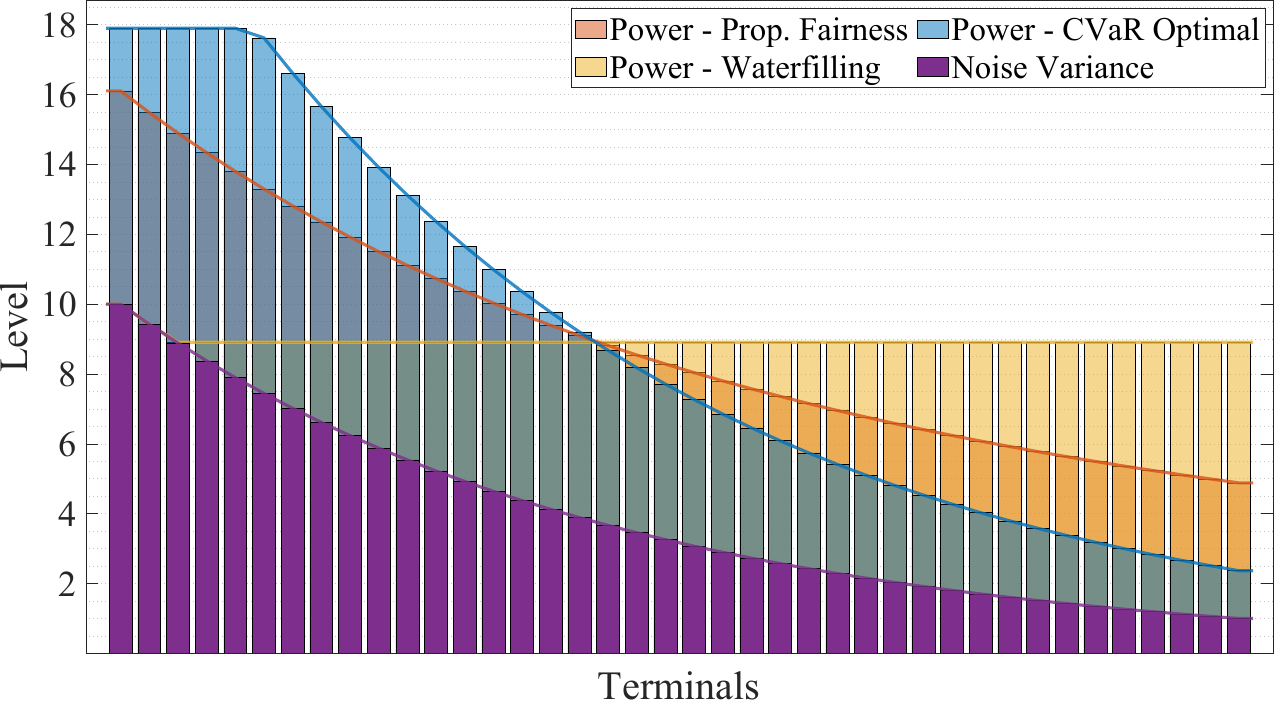}
\caption{Collective levels of allocated policies of logspaced terminals in $[1,10]$ under Edge Waterfilling and benchmark schemes.}
\label{fig:n40_bar}
\end{figure}
\begin{table}[t!]
\vspace{2pt}
\vspace{-.5\baselineskip}
\centering
\caption{SL$\alpha$Q values of distributed by $\mathcal{U}[1,10]$ (Scenario 1) and logspaced terminals in $[1,10]$ (Scenario 2) under Edge Waterfilling and the benchmark schemes at convergence.\vspace{-6pt}}
\footnotesize
\begin{tabular}{c||c|c|c}
\hline
Scenario & Prop. Fairness & Waterfilling & Edge Waterfilling\\ \hline
Scenario 1 & $0.5986$ & $0.5004$ & $0.6421$ \\ \hline
Scenario 2 & $0.6915$ & $0.4540$ & $0.8215$ \\ \hline
\end{tabular}
\vspace{9pt}
\label{table:slaq}
\end{table}

We also observe that the edge waterfilling algorithm adheres to precise waterfilling characteristics beyond the statistical cutoff, as shown in Figs.~\ref{fig:n100_bar} and \ref{fig:n40_bar}. However, the channel-opportunistic characteristics of classical waterfilling dissipate towards the proximal terminals, as the distributed power is gradually shifted towards the cell-edge users. For lower confidence levels $\alpha$, the \textit{robustification} towards the cell-edge terminals leads to stricter optimization of the sum-quantile rates of cell-edge terminals, resulting in a lower statistical bound on transmission rates through decreasing $t^*$ and provides robust and consistent transmission links. Further, CVaR-optimal policies are even stricter than proportionally fair allocation at the proximal terminals, prioritizing balance among terminals by explicitly preventing over-allocation towards high-quality transmission links. On the other extent, as the CVaR confidence level $\alpha$ increases, the statistical cutoff on noise variance decreases with a rising $t^*$, allowing greater variability in transmission rates across noise levels. In the limiting case as $\alpha \to 1$, the CVaR-optimal policy recovers classical waterfilling, and the SL$\alpha$Q objective reduces to the standard sumrate maximization problem. Lastly, the proposed Edge Waterfilling algorithm demonstrates strong scalability, making it effective and practical for large-scale network deployments (see Figs.~\ref{fig:n100_rates} and \ref{fig:n100_bar}). 

\begin{figure}[t!]
\vspace{-2pt}
\includegraphics[width=1.05\linewidth]{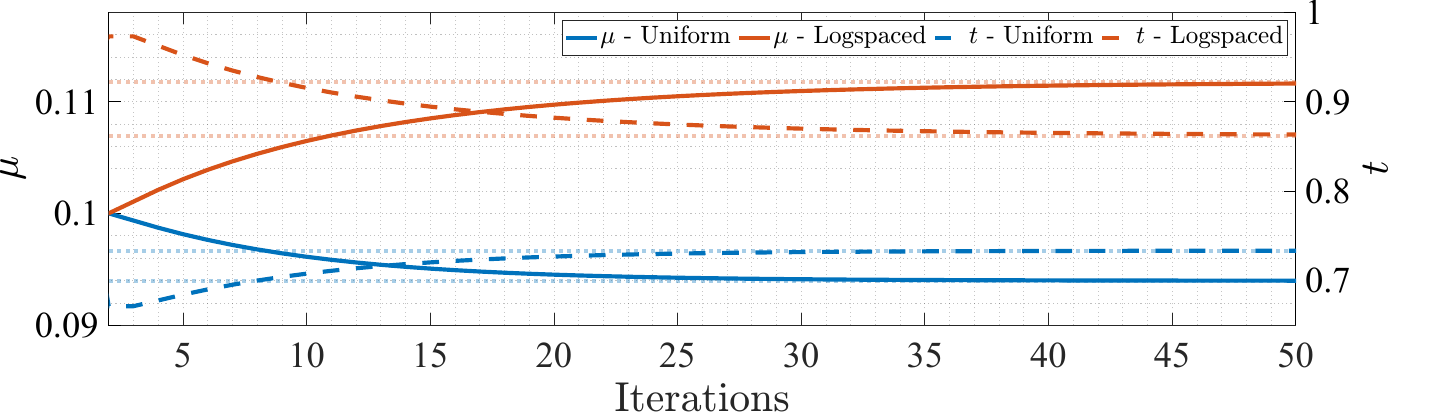}
\caption{$\mu$ and $t$ iterates in both scenarios.}
\label{fig:mu_t_iterates}
\vspace{4pt}
\end{figure}

\vspace{-9pt}
\section{Conclusion}
\vspace{-7pt}
\label{sec:conclusion}


We revisited the sum-least-$\alpha$th-quantile (SL$\alpha$Q) problem and proposed a risk-averse reformulation using Conditional Value-at-Risk to address deterministic resource allocation in point-to-point, multi-terminal, interference-free networks. The CVaR framework robustly and exactly maximizes the sum rate of the weakest $\alpha$-quantile terminals, in fact generalizing SL$\alpha$Q. Using Lagrangian duality, we presented closed-form primal solutions and an efficient and scalable inexact dual subgradient descent scheme, called Edge Waterfilling, together with a complete convergence analysis ensuring decaying optimality violation. Our numerical results confirmed the effectiveness of edge waterfilling in comparison to standard benchmark policies, even in large scale settings.


\vspace{-.5\baselineskip}

{\small \bibliography{Files/references}}

\end{document}